\documentclass[3p,times,twocolumn]{elsarticle}

\usepackage{ecrc}

\volume{00}

\firstpage{1}

\journalname{Nuclear and Particle Physics Proceedings}

\runauth{}

\jid{nppp}

\jnltitlelogo{Nuclear and Particle Physics Proceedings}

\usepackage{amssymb}

\usepackage[figuresright]{rotating}

\usepackage{aas_macros}

\begin{document}

\begin{frontmatter}



\dochead{}

\title{Ancient Pulsar Wind Nebulae as a natural explanation for unidentified gamma-ray sources}


\author{S. Kaufmann, O. Tibolla}

\address{Mesoamerican Center for Theoretical Physics (MCTP), Universidad autonoma de Chiapas (UNACH), Tuxtla Gutierrez, Mexico}

\begin{abstract}
A large part of the Galactic sources emitting very high energy (VHE; $> 10^{11}$  eV) gamma-rays are currently still unidentified. The evolution of Pulsar Wind Nebulae (PWNe) plays a crucial role in interpreting these sources. The time-dependent modeling of PWNe has been tested on a sample of well-known young and intermediate age PWNe; and it is currently applied to the full-sample of unidentified VHE Galactic sources. The consequences of this interpretation go far beyond the interpretation of ``dark sources'' (i.e. VHE gamma-ray sources without lower energies, radio or X-ray, counterparts): e.g. there could be strong implication in the origin of cosmic rays and (when considering a leptonic origin of the gamma-ray signal) they can be important for reinterpreting the detection of starburst galaxies in the TeV gamma-ray band. 
Moreover, the number of Galactic VHE sources is currently increasing with further observation by Imaging Atmospheric Cherenkov Telescopes (IACTs) and by the advent of more sensitive water Cherenkov telescopes such as HAWC (High-Altitude Water Cherenkov Observatory); therefore the physical interpretation of unidentified sources becomes more and more crucial.
\end{abstract}

\begin{keyword}
Cosmic Rays \sep Unidentified gamma-ray sources \sep Pulsar Wind Nebulae \sep HESS J1708-410 \sep HESS J1420-607 \sep IGR J1849-0000

\end{keyword}

\end{frontmatter}



\section{Introduction}

Pulsar Wind Nebulae (PWNe) with their highly efficient acceleration of particles to ultra-relativistic energies, have been studied since decades.  

With the advent of imaging atmospheric Cherenkov telescopes (IACT), the number of Galactic sources emitting at  TeV gamma-ray energies, raised to a total of around 130\footnote{TeVCAT (http://tevcat.uchicago.edu/) for a continously updated list}. Many of these sources are interpreted to be Pulsar Wind Nebula, but the largest part is still unidentified. 

Although much progress has been made in understanding the emission processes of the PWNe, e.g. the evidence for the accumulation of very high energy electrons in a PWN, such as HESS J1825-137 \cite{Aharonian2006_1825}, many questions remain.
Different morphology and brightness at TeV gamma-rays and lower energies give important insight into the interpretation of the emission processes.
It was suggested, that an evolved PWN can lead to a fairly bright gamma-ray source without any lower energy counterpart (e.g. \cite{HESSColl2011_1507Tibolla},\cite{deJager2009_ICRC},\cite{Tibolla2011_ICRC}]). 
Especially the discovery of the very high energy (VHE,E$>10^{11} eV$) source HESS J1507-622, off-set from the Galactic plane without reasonable low energy counterparts, was important, since an ancient PWN model is the only reasonable scenario which can physically describe the VHE emission of this unique source (e.g. \cite{HESSColl2011_1507Tibolla},\cite{Tibolla2011_ICRC},\cite{Vorster2013}).

The idea that most of the unidentified Galactic TeV gamma-ray sources could be PWNe started to collect more and more proofs.

\newpage

\section{Time-dependent modeling of PWNe and the complete sample}

After some earlier models  (e.g. \cite{HESSColl2011_1507Tibolla},\cite{deJager2009_ICRC},\cite{Tibolla2011_ICRC}]), we developed a leaky-box \cite{zhang} PWN model \cite{gamma12} and finally a new spatially-independent model to calculate the temporal evolution of the electron/positron spectrum in a spherically expanding PWN \cite{Vorster2013}. 
As also described in \cite{Vorster2013}, the magnetic field is decreasing with time ($B \propto t^{-1.3}$) as theoretical models predict \cite{chev84}. 
A consequence of this model is, that during the evolution of the PWN, the synchrotron luminosity fades below the sensitivity of the current generation of X-ray satellites in short time scales, while the TeV gamma-ray emission remains bright and continues to increase thanks to the measured accumulation of leptons.

This new code has been tested \cite{Vorster2013} on the young PWN G21.5-0.9 (HESS J1833-105), and successfully applied to the two unidentified VHE gamma-ray sources HESS J1427-608 and HESS J1507-622 \cite{Vorster2013}, strengthening the argument that the unidentified VHE gamma-ray sources can indeed be identified as aged PWNe. 
We tested our model with young, ``middle-age'' and very old PWNe, increasing the number of tested sources, i.e. successfully modeling and classifying four unidentified sources as PWNe (HESS J1837-069, HESS J1616-508, HESS J1702-420, HESS J1708-410)\cite{Tibolla2013_ICRC}, while another unidentified source HESS J1804-216 could not be described by our model (and indeed there are several indications which would suggest this source to be a shell-type SNR) \cite{Tibolla2013_ICRC}.

\begin{table}
\begin{tabular}{|c|c|}
\hline
\multicolumn{2}{|c|}{unidentified VHE $\gamma$-ray sources}  \\
\hline
MAGIC J0223+403			&	HESS J1843-033	\\
ARGO J0409-0627			&	0FGL J1844.1-0335	\\
2HWC J0819+157			&	2HWC J1844-032	\\
2HWC J1040+308			&	HESS J1844-030	\\
2HWC J1309-054			&	HESS J1848-018	\\
HESS J1457-593			&	2HWC J1852+013 \\
HESS J1503-582			&	HESS J1852-000	\\
HESS J1614-518			&	HESS J1857+026	\\
HESS J1626-490			&	MAGIC J1857.6+0297	\\
HESS J1634-472			&	HESS J1858+020	\\
HESS J1641-463			&	0FGL J1900.0+0356	\\
HESS J1646-458 (Wd1?)		&	2HWC J1902+048	\\
HESS J1729-345			&	1HWC J1904+080c	\\
HESS J1741-302			&	2HWC J1907+084	\\
HESS J1745-303* (hot spot C)	&	ARGO J1910+0720 \\
VER J1746-289			&	HESS J1912+101	\\
HESS J1746-285			&	2HWC J1914+117	\\
MAGIC J1746.4-2853		&	2HWC J1921+131	\\
HESS J1747-248 (Terzan 5?)	&	2HWC J1928+177	\\
HESS J1804-216			&	2HWC J1938+238 \\
HESS J1808-204			&	HESS J1943+213	\\
HESS J1809-193			&	2HWC J1949+244	\\
HESS J1813-126			&	2HWC J1953+294	\\
2HWC J1825-134			&	2HWC J1955+285	\\
HESS J1826-130			&	2HWC J2006+341	\\
HESS J1828-099			&	VER J2016+371	\\
2HWC J1829+070			&	VER J2019+368 \\
HESS J1832-085			&	MGRO J2019+37	\\
HESS J1834-087			&	TeV J2020+380d	\\
2HWC J1837-065			&	VER J2019+407	\\
HESS J1841-055			&	MGRO J2031+41 \\
1HWC J1842-046c			&	TeV J2032+4130\\
\hline
\end{tabular}
\caption{Sample of unidentified VHE $\gamma$-ray sources as described in the text. *for more details about this morphologically complex source \cite{Aharonian2008}}
\end{table}

Anyway it is necessary to apply the model to a larger number of known PWNe in order to derive a statistically significant set of parameters that can be used as a guideline for the PWN evolution; moreover, this additional test allows us to constrain better the unidentified VHE gamma-ray sources in the frame of PWN evolution and, eventually, to discard the unidentified sources that do not fit in this picture. 
Some further examples of this on-going work will be presented in the next section.

We continue updating the sample to achieve a complete sample of the TeV bright PWNe, the candidate PWNe and the unidentified VHE gamma-ray sources. Tables 1 and 2 summarize the Galactic TeV $\gamma$-ray sources interpreted as PWN (or PWN candidate) and the sources which are laking a reasonable low energy counterpart or a reasonable emission model and hence are still classified as unidentified $\gamma$-ray sources. 
The ``border line'' between candidate PWNe and unidentified sources is obviously rather faint (i.e. the candidates are by definition not firmly identified): therefore here we consider as candidate PWNe only the sources that have strong indications of their PWN nature and no other likely scenario to explain their VHE gamma-ray emission.
Please note, that the Galactic center sources are not included in our list (with the exception of the sources in this region that are clearly detached from the Galactic center itself, such as HESSJ1741-302 \cite{Tibolla2008}, HESSJ1745-303 \cite{Aharonian2008}), for the complexity of the region and the dominance of diffuse emission, even if several models (especially at lower energies) predict a strong leptonic component in that region.
Moreover, it is interesting to note that some of the candidate PWNe shown in Table 2, as well as some unidentified sources, are also coincident with SNR shells and/or with molecular clouds: in these cases, the application of our model is even more important in order to constrain the source identification and disentangle between different scenarios.

\begin{table}
\begin{tabular}{|c|c|}
\hline
\multicolumn{2}{|c|}{PWNe (and PWN candidates)}  \\
\hline
CTA 1			& HESS J1702-420	\\		
3C 58			& PSR B1706-44	\\
Crab			& HESS J1708-410	\\
LHA 120-N 157B		& HESS J1708-443	\\
0FGL J0631.8+1034	& HESS J1718-385	\\
Geminga			& HESS J1745-303* (hot spot B)	\\
Vela X			& HESS J1747-281	\\
HESS J1018-589B		& HESS J1809-193	\\
HESS J1026-582		& HESS J1813-178	\\
HESS J1119-614		& HESS J1818−154	\\
HESS J1303-631		& HESS J1825-137	\\
HESS J1356-645		& HESS J1831-098	\\
Kookaburra (PWN)	& HESS J1833-105	\\
Kookaburra (Rabbit)	& HESS J1837-069	\\
HESS J1427-608		& HESS J1846-029	\\
HESS J1458-608		& IGR J18490-0000	\\
HESS J1507-622		& HESS J1857+026	\\
MSH 15-52		& MGRO J1908+06	\\
SNR G327.1-01.1		& SNR G054.1+00.3	\\
HESS J1616-508		& 0FGL J1958.1+2848	\\
HESS J1632-478		& 0FGL J2021.5+4026  		\\
HESS J1640-465		& Boomerang \\
\hline
\end{tabular}
\caption{Sample of PWNe and PWN candidates as described in the text.  *for more details about this morphologically complex source \cite{Aharonian2008}}
\end{table}

Moreover, the number of Galactic VHE sources is currently increasing with further observation by Imaging Atmospheric Cherenkov Telescopes (IACT) and by the advent of more sensitive water Cherenkov telescopes such as HAWC (High-Altitude Water Cherenkov Observatory). 
The source catalog of the HAWC observatory \cite{Abeysekara2017_2HWC}, obtained during 17 month of observation already added many new unidentified VHE $\gamma$-ray sources.
Together with its advent of being an all-sky instrument, the HAWC observatory will also extend the TeV spectra to higher energies \cite{albertoCRBTSM}.

\begin{figure}
\centering
\includegraphics[width=\linewidth]{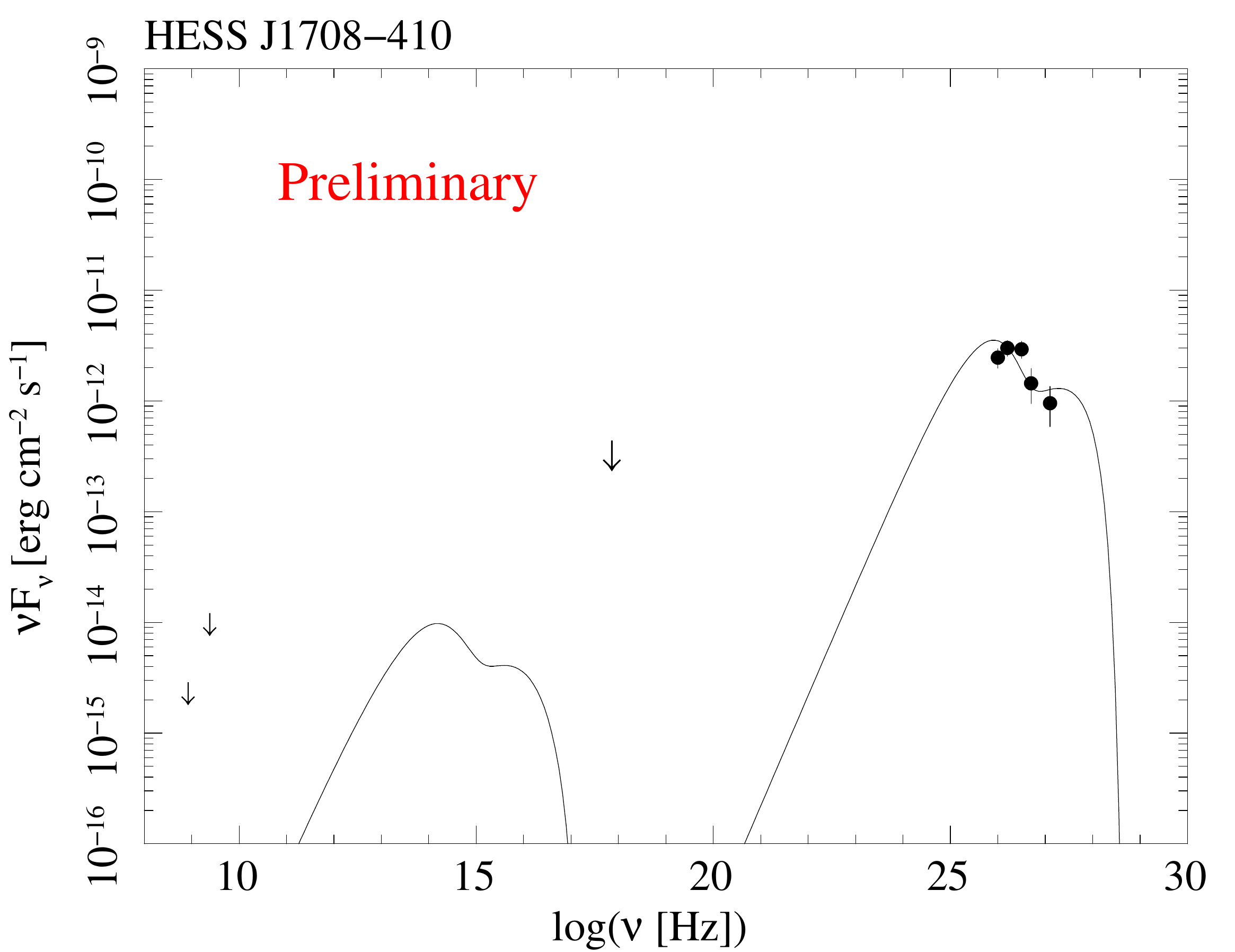}
\caption{Spectral energy distribution (SED) of HESS J1708-410 with the applied time-evolution PWN model. H.E.S.S. data points are taken from \cite{Aharonian2006_survey}; the radio (in the 843 MHz and 2.4 GHz bands) and X-ray (XMM-Newton) from \cite{VanEtten2009}.}
\label{SED1708}
\end{figure}

The model is applied to the PWNe, candidate PWNe and unidentified sources. To illustrate the changes in the spectral energy distribution, we present examples with preliminary results on the study of young, intermediate and old PWN system. 

\subsection{Old PWN system}

An aged PWN is an excellent tool to explain and describe so-called ``dark sources''.  
A prominent example of an unidentified VHE source which can only be described by invoking an ancient PWN system is HESS J1507-622 \cite{HESSColl2011_1507Tibolla} \cite{Vorster2013}. 
The lack of X-ray counterparts makes it difficult to clearly identify this source and the ancient PWN model seems the most reasonable and only plausible description. 
As determined in \cite{Vorster2013}, this ancient PWN system has a present day magnetic field of $B_{\rm{age}} \sim 1.7 \rm{\mu G}$ and an age of $24$ kyrs.
However different interpretations of HESS J1507-622 distance \cite{eger} can lead to even more extreme scenarios.

HESS J1708-410 \cite{Aharonian2008_unid} \cite{Aharonian2006_survey} is often cited among the ``dark sources'' (e.g. \cite{HESSColl2011_1507Tibolla},\cite{Tibolla2011_ICRC}), given the lack of X-ray and radio counterparts. 
Let us have a look into it in more detail. 
An old PWN system, with a highly suppressed synchrotron peak seems to be a possible description and this seems strengthened by the fact that no clear GeV gamma-ray counterpart could be found. 
A PWN with an age of approx. $8000$ years and a present day magnetic field of $B_{\rm{age}} \sim 0.2 \rm{\mu G}$ (due to this small value, this source has a low synchrotron luminosity, while it remains bright at TeV energies) is a plausible description of the detected TeV spectrum and the low energy upper limits (see Fig. \ref{SED1708}). 

\begin{figure}
\centering
\includegraphics[width=\linewidth]{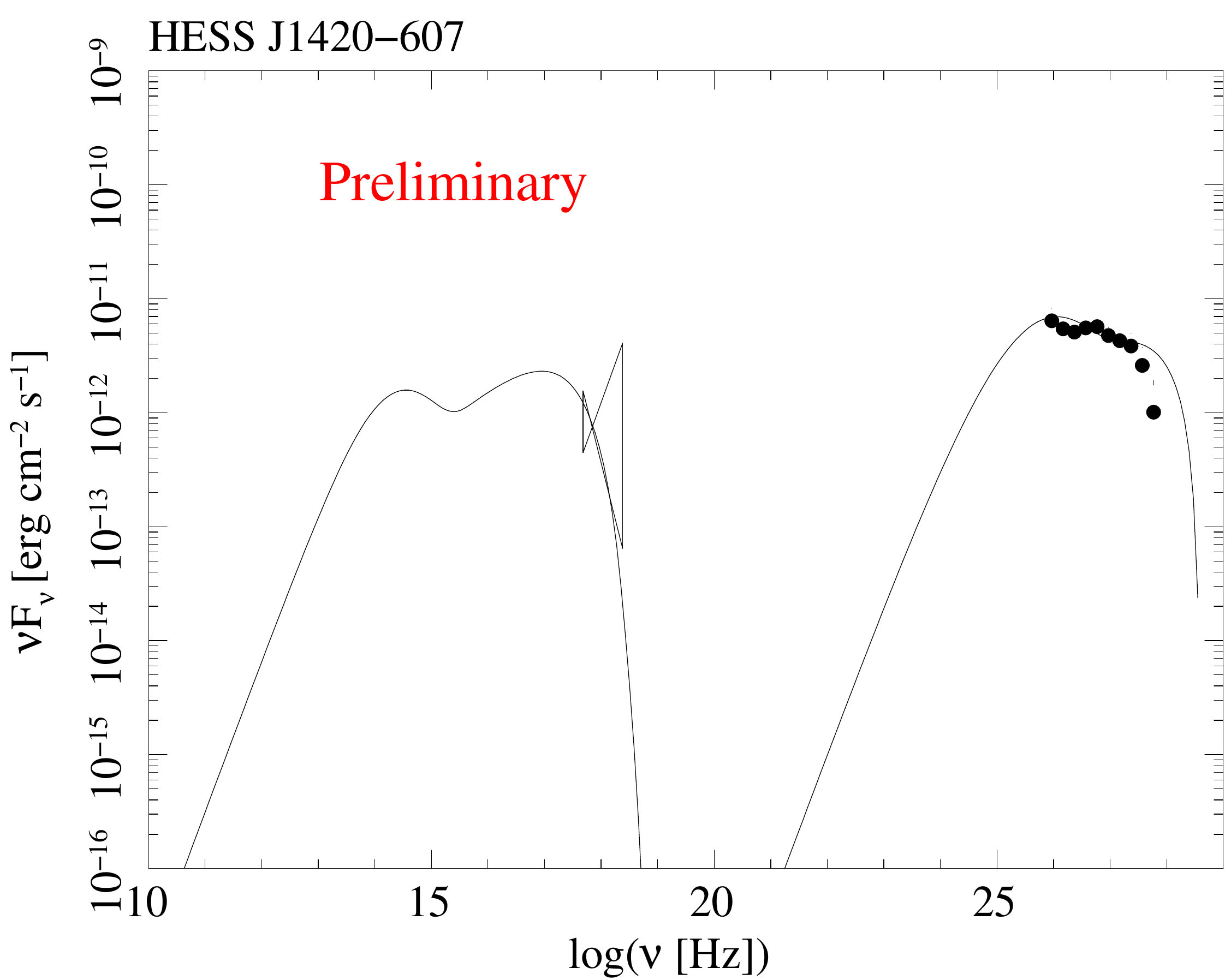}
\caption{SED of HESS J1420-607  with the applied time evolution PWN model. 
X-ray spectrum taken from \cite{Ng2005} and the VHE $\gamma$-ray spectrum from \cite{Aharonian2006_Kookaburra}.
}
\label{SED1420}
\end{figure}

\subsection{Young and intermediate age PWN system}

The Kookabura PWN (HESS J1420-607,\cite{Aharonian2006_Kookaburra}) is a prominent example of young PWN system. 
HESS J1420-607 has been classified as a PWN \cite{Acero2013} and is spatially coincident with the pulsar PSR J1420-6048 ($\dot{E} = 1 \times 10^{37} \; \rm{erg \;  s^{-1}}$, characteristic age = 13 kyr, distance = 5.6 kpc). 
The GeV emission detected in the 3FGL catalog is dominated by this gamma-ray pulsar. 
X-ray studies have been performed by \cite{Ng2005} and faint diffuse emission, surrounding the pulsar, was found. 
Therefore this PWN system is very well constrained and so we can apply the PWN model (see Fig. \ref{SED1420}), considering the characteristics of the Pulsar, yield in a present day magnetic field of  $B_{\rm{age}} \sim 20\; \rm{\mu G}$ and an age of approx. $2000$ years. 

\begin{figure}
\centering
\includegraphics[width=\linewidth]{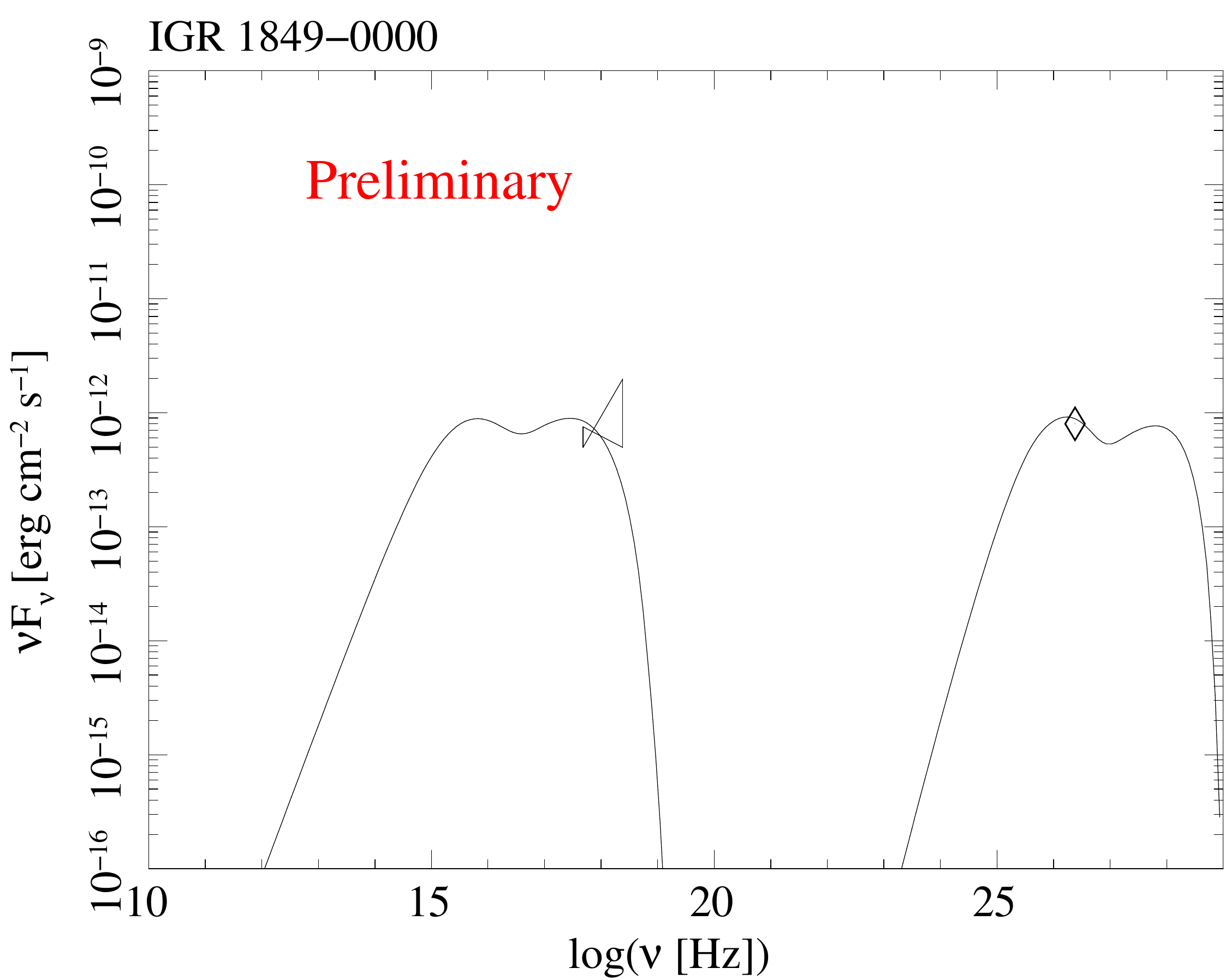}
\caption{SED  of IGR J1849-0000 with the applied PWN model. 
Very high energy gamma-ray flux by \cite{Terrier2008} and XMM-Newton spectrum taken from \cite{Laila}.}
\label{SEDIGR}
\end{figure}

IGR J1849-0000 can be interpreted as an intermediate age system. 
Very high energy gamma-rays have been detected by \cite{Terrier2008}, and studies at X-ray energies with RXTE revealed the detection of the X-ray pulsar PSR J1849-001  ($\dot{E} = 9.8 \times 10^{36} \; \rm{erg \; s^{-1}}$, characteristic age = 42.9 kyr) with a pulsation period of 38.5 ms \cite{Gotthelf2011}. 
Further studies with XMM-Newton show a faint emission surrounding the pulsar which can indeed be interpreted as PWN \cite{Laila}.
Due to the clear connection of the pulsar with the PWN, also in this case the characteristics of the pulsar have been used for efficiently constraining the time evolution PWN model.  
A present day magnetic field of $B_{\rm{age}} \sim 3\; \rm{\mu G}$ and an age of approx. $3000$ years for the PWN result in a good description of the X-ray and TeV gamma-ray emission (see Fig. \ref{SEDIGR}).

\section{Conclusions}


Even if the results shown here are still preliminary, the time-dependent model (described in \cite{Vorster2013}) is clearly a powerful tool to study PWNe, but not only; in fact, once more, it strengthen the idea 
that most of the unidentified VHE gamma-ray sources can be described within a PWN framework and this will more than double (as underlined in Table 1 and 2) the population of VHE PWNe. 
In particular it can confirm the hypothesis, suggested by \cite{HESSColl2011_1507Tibolla} and \cite{deJager2009_ICRC}, that the so-called ``dark sources'' are indeed ancient PWNe.
Moreover strengthening the idea about the very long lifetime of inverse Compton emitting electrons in VHE gamma-ray PWNe could have strong implications, far beyond the interpretation of ``dark sources''. 
It could, for example, have implications also about the origin of Cosmic rays. 
Moreover, the dominant population of identified sources in the VHE sky is not represented by shell-type SNRs, but PWNe are by far the most numerous. 
PWNe are clearly an efficient and established machine to accelerate leptons. 
And, even more, you can find in literature tentatives to see if at the termination shock of the pulsar wind, also hadrons could be accelerated as well as leptons (e.g. \cite{Bednarek1997}, \cite{Atoyan1996}, \cite{Cheng1990}, \cite{Bednarek2003}). 
Moreover it has been underlined by \cite{Mannheim2012} that PWNe can be important for reinterpreting the detection of starburst galaxies in the TeV gamma-ray band.

\paragraph*{Acknowledgments}

We acknowledge the grants Conacyt CB-258865 and Royal Society Newton Advanced Fellowship-180385.




\nocite{*}
\bibliographystyle{elsarticle-num}
\bibliography{PWNreferences_short}

\end{document}